\documentclass[ACS,STIX1COL]{WileyNJD-v2}
\articletype{ORIGINAL ARTICLE}

\usepackage{mathtools}
\mathtoolsset{showonlyrefs=true}

\received{xxx}
\revised{xxx}
\accepted{xxx}

\newcommand{\onlinecite}[1]{\hspace{-1 ex} \nocite{#1}\citenum{#1}} 

\usepackage{bm}

\begin{document}

\title{
Yuri L'vovich Klimontovich, his theory of fluctuations and its impact on the kinetic theory
}

\author[1] {Michael Bonitz}
\author[2] {Anatoly Zagorodny} 

\authormark{M. Bonitz  \textsc{et al}}

\address[1]{\orgdiv{Institut f\"ur Theoretische Physik und Astrophysik}, \orgname{Christian-Albrechts-Universit\"at zu Kiel}, \orgaddress{\state{Leibnizstra{\ss}e 15, 24098 Kiel}, \country{Germany}}}

\address[2]{\orgdiv{Bogolyubov Institute for Theoretical Physics}.
\orgname{National Academy of Sciences of Ukraine},

\orgaddress{\state{Metrolohichna Str.14-b, 03143 Kyiv}, \country{Ukraine}}}

\corres{*\email{bonitz@physik.uni-kiel.de}}

\abstract{Yuri L'vovich Klimontovich (28.09.1924–26.10.2002) was an outstanding theoretical physicist who made major contributions to kinetic theory. On the occasion of his 100th birthday we recall his main scientific achievements.
}


\maketitle


\section{Introduction}\label{s:introduction}
Yuri L'vovich Klimontovich belongs to the cohort of outstanding theoretical physicists of the 20th century who made a fundamental contribution to the progress in the kinetic theory of many-particle systems and statistical physics. Without his fundamental papers and books, see Refs.~\onlinecite{klimontovich_67}--\onlinecite{klimontovich-open3}, it is difficult  to imagine many fields of theoretical physics (such as statistical plasma theory, kinetic description of nonequilibrium processes in non-ideal gases and nonideal plasmas, theory of turbulence, theory of open systems, fluctuations in quantum and relativistic systems, etc.). His most important achievement is the theory of fluctuations which he constructed on the basis of his original microscopic phase space density, see Sec.~\ref{s:classical_fluctuations}. This concept was very influential in plasma physics and kinetic theory in the Soviet Union and it was also being picked up by many theorists in Western Europe and the U.S. After Yu.L. Klimontovich passed away in 2002 his scientific achievements were discussed in a series of papers, complemented by personal recollections, e.g. \onlinecite{klimontovich-80years,bonitz_cpp_03,balescu_jpcs_05}, more articles are listed in Ref.~\onlinecite{bonitz_cpp_03}. In 2004 a workshop ``Physics of nonideal plasmas'' in honor of Klimontovich took place in Kiel, Germany, organized by one of the authors (MB). The conference proceedings, Ref.~\onlinecite{jpcs2005} contain  25 articles written by well-known scientists in his honor.

20 years later, Klimontovich's scientific ideas are still of high interest, as can be seen by the articles in this special issue. The present authors were both students of Yu.L. Klimontovich and share their personal views on his scientific results in this article.

\section{Personal life}\label{s:life}
\begin{figure}
    \centering
    \includegraphics[width=0.5\textwidth]{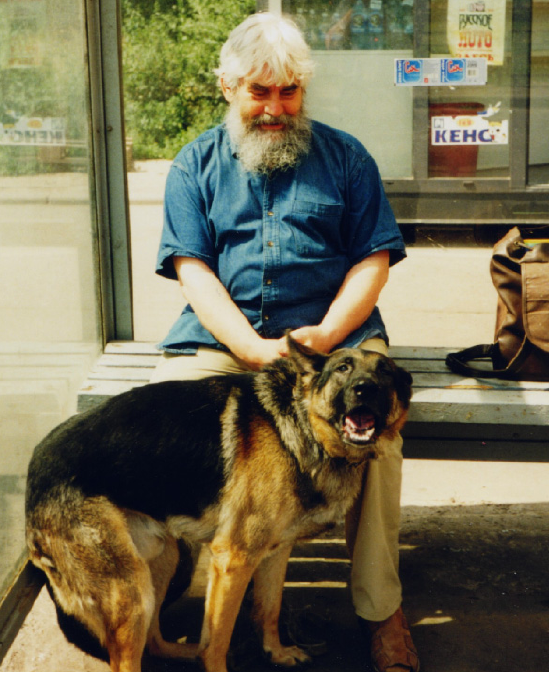}
    \caption{Yuri Lvovich Klimontovich in Moscow in 1999. Photo by M.~Bonitz}
    \label{fig:klim-photo}
\end{figure}
Yu.L.~Klimontovich was born on September 28, 1924, in Moscow. His parents belonged to noble families which fact has turned out to be a sufficient reason for his father to be arrested under Stalin's regime in 1931 and killed only two weeks later [\onlinecite{bonitz_cpp_03}]. Similarly, Joseph Maevsky, the father of Klimontovich's wife Svetlana Maevska, was arrested and shot in December 1937. So both families suffered a great tragedy and this also influenced Yuri Lvovich's views throughout his life. He was always critical towards the Soviet system which could be felt in personal discussions we had with him.

Nevertheless, Yuri managed to enter the Lomonosov Moscow State University and to graduate from it in 1948. After graduation, Yu.L.~Klimontovich continued his scientific activity as a Ph.D. student of Prof. N.N.~Bogolyubov -- a famous mathematician and theoretical physicist, well-known for his contributions to particle physics, Bose condensation, superfluidity, and also one of the founders of the modern theory of kinetic processes (BBGKY-hierarchy). Bogolyubov’s outstanding ideas concerning the dynamical problems of statistical mechanics have had a profound  influence on Klimontovich and some he took up in his further scientific activities. In 1951 Yu.L.~Klimontovich successfully defended his candidate (PhD) thesis and started to work out his own fundamental scientific ideas concerning the microscopic description of many-particle systems, in particular, nonideal nonequilibrium gases and plasmas.

\section{Microscopic phase space distribution}\label{s:classical_fluctuations}
The key point of the theory developed by Yu.L.~Klimontovich [\onlinecite{klimontovich_jetp_57}] is the description of a microscopic state of the system under consideration in terms of the microscopic phase space density of a classical $N$-particle system
\begin{align}
    N(X,t) = \sum_{i=1}^N \delta[X-X_i(t)]\,,\quad X=(\textbf{r},\textbf{p})\,,
\end{align}
where $X$ is a point in six-dimensional phase space and $X_i(t)$ are the trajectories of all particles. This function is a natural generalization of the charge density of electrodynamics. The function $N$ satisfies a continuity equation in the phase space which in the present case has the form of the well-known Vlasov equation. However, it differs essentially from the Vlasov equation for the well-known distribution function $f$ since the physical as well as mathematical meaning of $N(X, t)$ is fundamentally different. 
In fact, $f$ satisfies the Vlasov equation only if collisions are neglected. In contrast, the  equation for $N$ (which is called ``Klimontovich equation'' in many text books and articles), Eq.~(\ref{eq:EOM:N}), is exact: 
\begin{equation}
    \left[ \partial_t+\mathbf{v}\cdot\nabla_\mathbf{r}+\mathbf{F}^\mathrm{M}(\mathbf{r},t)\cdot\nabla_\mathbf{p} \right]N(X,t)=0\,, \label{eq:EOM:N}
\end{equation}
where we introduced the microscopic force $\mathbf{F}^\mathrm{M}\coloneqq -\nabla_\mathbf{r}[V+ U^\mathrm{M}]$ defined in terms of the gradient of the external potential $V$ and the microscopic mean-field potential (microscopic version of the Vlasov or Hartree potential) $U^\mathrm{M}$ given by
\begin{equation}
   U^\mathrm{M}(\mathbf{r},t)\coloneqq \int W(\mathbf{r},\mathbf{r}')N(X',t)\,\mathrm{d}X'\,,
   \label{eq:u-def}
\end{equation}
where $W$ is the pair interaction between the particles. In the case of a plasma, $\textbf{F}^{\rm M}$ is the microscopic Lorentz force that couples the particles to the dynamics of the electromagnetic field.
The key difference to the standard Vlasov equation is that $N$ is a random function that depends on random initial conditions of all particles. An ensemble average $\langle \dots \rangle$ yields the distribution function $f$,
\begin{align}
    N(X,t) = \langle N(X,t) \rangle +\delta N(X,t) \equiv n f(X,t) + \delta N(X,t)\,,
    \label{eq:n-def}
\end{align}
but $N$, in addition contains fluctuations around the mean. 
The Vlasov equation for $f$ is recovered from an ensemble average of Eq.~\eqref{eq:EOM:N}, but this equation contains an additional term proportional to the average of the product of two random functions,
\begin{align}
    I(X,t) \sim \langle \delta \textbf{F}^{\rm M}\nabla_\textbf{p}\delta N\rangle \,,
\end{align}
which is, of course, the collision integral of the kinetic equation (neglected in the standard Vlasov equation for $f$).
Thus, Klimontovich found a very original way to express collisions (correlations) in terms of particle and field fluctuations. The equation of motion for the fluctuations, $\delta N$, follows immediately by computing the difference of Eq.~\eqref{eq:EOM:N} and its ensemble average. 
As is seen from the definition of $N(X, t)$, Eq.~\eqref{eq:n-def}, the solution of this equation requires the knowledge of the phase space trajectories of all particles and, therefore,  the problem is equivalent to the full $N$-body problem. However, the formulation of the microscopic description in terms of $N(X, t)$ is much more convenient for the development of many important approximations of kinetic theory and approximation schemes. Yu.L.~Klimontovich showed how to fully exploit this advantage and developed a completely new treatment of the theory of nonequilibrium plasmas and fluctuation electrodynamics in particular. The proposed approach turned out to be especially useful for the derivation of the collision terms and various generalizations thereof (including retardation, electromagnetic effects, etc.), for more details, see Ref.~[\onlinecite{schroedter_cpp_24}]. 

During the development of this new approach Yu.L.~Klimontovich fruitfully collaborated with V.P.~Silin. This collaboration resulted in the theory of excitations in a quantum electron gas (which was developed before the theory of Lindhard, which it contains as limiting case) [\onlinecite{klimontovich_dan_52, klimontovich_jetp_52}] and microscopic quantum description of Coulomb systems. 
In 1962 Yu.L.~Klimontovich received his second scientific degree (doctor of physics and mathematics) from the Steklov Mathematical Institute. From 1964 and until his last days he occupied held a professorship at Lomonosov Moscow State University. In the same year 1964 he published his first book [\onlinecite{klimontovich_67}] which opened a new page in the kinetic description of nonequilibrium plasmas. After the publication of this book, the terms ``Klimontovich-Silin approach'', ``Klimontovich distribution function'' and ``Klimontovich equation'' became widely used in the scientific literature of gases and plasmas, in particular, in the Soviet Union.
While most of the work of Klimontovich was devoted to classical systems he also applied his approach to quantum systems, e.g.~[\onlinecite{klimontovich_ufn_60}]. He was well aware of the methods of second quantization and also the theory of equilibrium and nonequilibrium Green functions that are nowadays the common method in quantum many-body physics, quantum electrodynamics, and condensed matter physics. The close relation of Klimontovich's approach of microscopic phase space density and the approach of second quantization has been established in Ref.~[\onlinecite{dufty_jpcs_05}].

No doubts, Klimontovich’s approach was an important contribution to the -- at that time -- very strong ``Soviet'' fluctuation theory (from radio physics, laser theory, plasma physics, kinetic theory, turbulence etc.) which included well-known proponents such as A.I.~Akhiezer, B.B.~Kadomtsev, M.A.~Leontovich, P.S.~Landa, A.A.~Rukhadze, S.M.~Rytov, V.P.~Silin, A.G.~Sitenko, R.L.~Stratonovich, V.N.~Tsytovich, I.P.~Yakimenko and their numerous students and co-workers.

\begin{figure}
    \centering
    \includegraphics[width=0.95\textwidth]{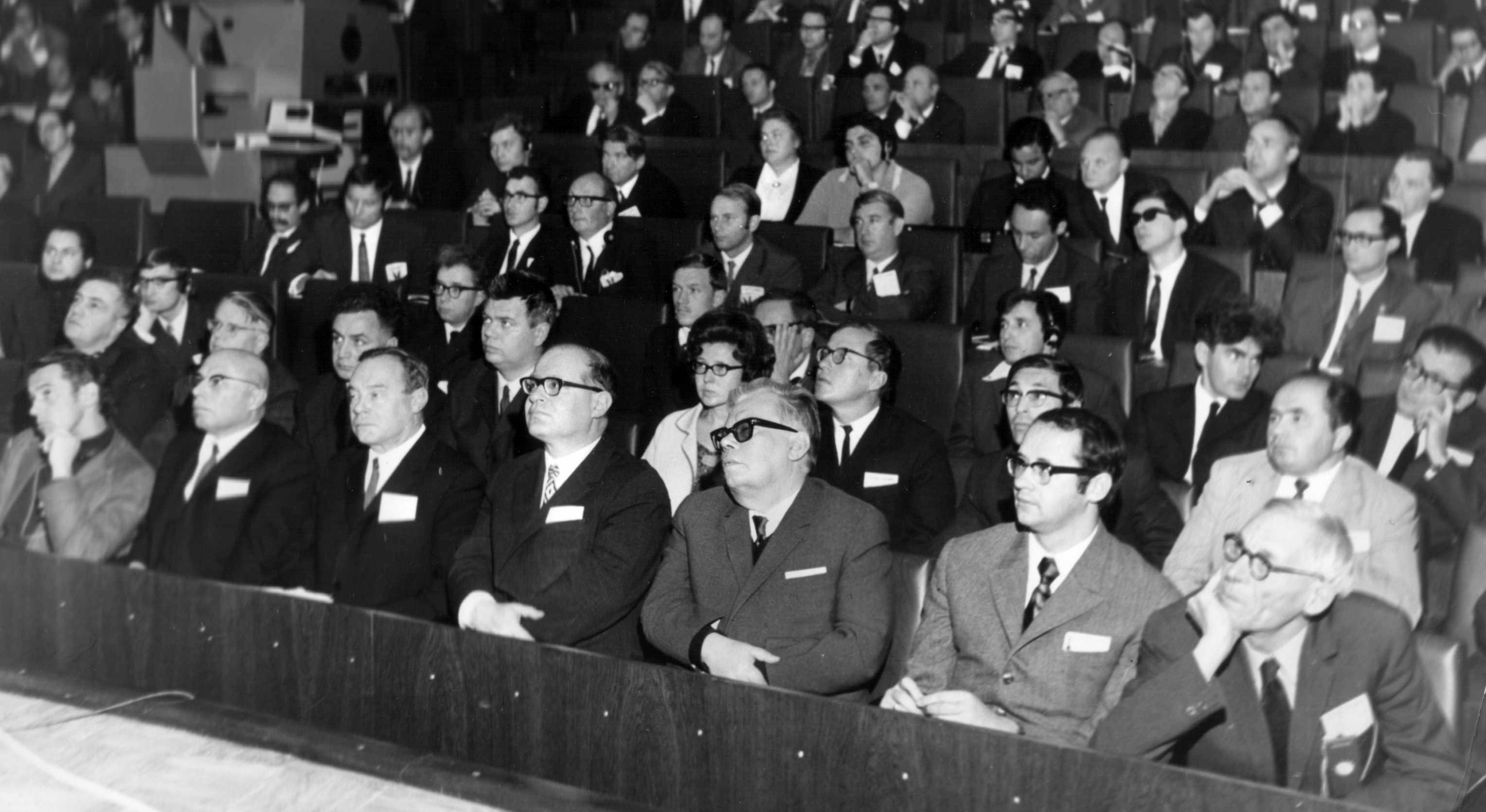}
    \caption{Participants of the plenary session of the International (Kyiv) Conference on Plasma Theory, Kyiv, 1971, including -- first row: M.~Leontovych (right), N.N.~Bogolyubov (third from the right), A.S.~Davydov (second from the left); second row:  B.~Kadomtsev,  first from the right. Third row: A.~Sitenko and  Yu.~Klimontovich (first and second from right); fourth row: D.~Kremp, G.~Kelbg, W.~Ebeling, and W.D.~Kraeft (5th-8th from right). Photo by S.~Filipchuk}
    \label{fig:kiev-conf-71}
\end{figure}
\section{Later applications of Klimontovich's approach}\label{s:later}
Another advantage of the microscopic phase space density formalism is the possibility to directly extend it to the case of nonideal gases and nonideal plasma, as well as to the description of large-scale (kinetic) fluctuations [\onlinecite{klimontovich_75}]--[\onlinecite{klimontovich_82}]. With the microscopic equations being smoothened over what he called a ``physically infinitesimal'' time or spatial scale,  $\tau_{\mathrm{ph}}$  or   $l_{\mathrm{ph}}$, one obtains an equation for the smoothened microscopic phase density. This equation has the form of a kinetic equation with the collision term determined by the microscopic fluctuation spectra (i.e. fluctuations with correlation times much smaller than $\tau_{\mathrm{ph}}$ that defines the level of kinetic description). However, the smoothened microscopic phase space density still can be treated as a random quantity (since the smoothening is performed over an infinitesimal time interval only) and thus the kinetic equation can be used as a basic equation to work out the theory of large-scale fluctuations (in the same way as the theory of microscopic fluctuations). Large-scale fluctuation spectra generate additional collision terms responsible for the relaxation of random perturbations of distribution functions (for example, in the case of turbulent systems). Yu.L.Klimontovich has developed this general approach and, besides that, has shown how to practically introduce the relevant scales $\tau_{\mathrm{ph}}$ and $l_{\mathrm{ph}}$   for various systems and thus has opened new possibilities for the description of kinetic fluctuations in both classical and quantum systems. Moreover, the introduction of appropriate scales, $\tau_{\mathrm{ph}}$ or $l_{\mathrm{ph}}$  makes it possible to find a direct way of unifying the kinetic and gas-dynamic and hydrodynamic descriptions. Indeed, Klimontovich used this method to investigate fluctuations and spectral properties in a huge class of systems, e.g. [\onlinecite{klimontovich_ufn_83, klimontovich_ufn_87}], and many of these results still today hold substantial potential for application in many fields. From the mathematical side, Yuri Klimontovich was convinced that any dynamics in many-particle systems is necessarily dissipative (non-Hamiltonian), and he introduced dissipation by averaging the relevant equations (such as the Liouville equation or the Schrödinger equation) over the scales $\tau_{\mathrm{ph}}$ and   $l_{\mathrm{ph}}$. This gave rise to results that significantly differed from the commonly used expression for non-dissipative systems. A particularly striking example was his discussion of the fluctuation-dissipation relation (Nyquist's theorem, Callen-Welton formula) where he disputes the standard approach based on non-dissipative many-particle theory and
which he included in a number of articles and also his text book [\onlinecite{klimontovich_82}]. In 1986 he submitted a review paper on this subject to Uspekhi Fiz. Nauk [Sov. Physics Uspekhi], cf. Ref.~[\onlinecite{klimontovich_ufn_87}] which apparantly caused some stir in the editorial board after they had received negative referee reports. Given the high reputation of Klimontovich in the field of Statistical Physics and his activities as a university teacher they decided (the decision of the board is reprinted just before Klimontovich's article, Ref.~[\onlinecite{klimontovich_ufn_87}]) to publish his article side by side with two other articles that disagreed with Klimontovich. Moreover they did not give him the opportunity to answer this criticism and even stated that ``these question should not be further discussed in the journal''. Aside from this unusual editorial policy to terminate a scientific discussion, the content of these additional articles that were written by leading Soviet theoreticians, is interesting. The authors were  V.L.~Ginzburg and L.P.~Pitaevskii and by V.I.~Tatarski,  respectively, cf. Refs.~[\onlinecite{ginzburg_ufn_87,tatarskii_ufn_87}], and they presented strong arguments why the traditional quantum Nyquist formula is correct, but not the result of Klimontovich. Even today it is very interesting to compare both view points which might stimulate further research in the area of quantum many-body physics.

Yuri~Klimontovich also made important contributions to the theory of plasma-molecular systems [\onlinecite{zeiger_74,klimontovich_80,klimontovich_82,klimontovich_phys-rep_89}]. In this theory, the contribution of a molecular subsystem (atoms and molecules) to electromagnetic fluctuations and to the structural properties of the system under consideration is treated on equal footing with that of plasma particles. The kinetic theory of electromagnetic processes in molecular systems and plasmas that had been developed by Klimontovich earlier, has provided a natural basis for such a unification. In this point it is extremely important that Klimontovich’s approach (which is used as a basis in formulating the above mentioned kinetic theory) was developed for both classical and quantum models of particles and fields. This makes it possible to consistently describe electromagnetic field interaction with electrons, ions, and atoms, as well as the processes of ionization and recombination. Effects associated with radiation friction, finite line-widths of atomic orbitals and collisions between plasma particles and atoms can be also taken into account. The fluctuation theory of bremsstrahlung in plasma-molecular systems can be regarded another example of the efficiency of this powerful theory~[\onlinecite{zagorodny_cpp_89}].

During the about 25 final years of his life Yuri Klimontovich was fascinated by and involved with the statistical theory of open systems. He was interested in the most fundamental questions of this then still new field: what is a measure of stochasticity in an open system (in particular, under developed turbulence)? How to introduce such a measure? Which state is more regular (more ordered) -- equilibrium or a turbulent state? What are the general principles of selforganization? Yuri~Klimontovich devoted special attention to the problems of open and dissipative classical and quantum systems (as we already discussed above). In these fields he also managed to obtain interesting results, for an overview see Ref.~[\onlinecite{klimontovich_ufn_89}] and the text books ~[\onlinecite{klimontovich-open1}--\onlinecite{klimontovich-open3}]. These include the proof of the S-theorem [\onlinecite{klimontovich_zpb_87,klimontovich_zpb_88}], derivation of the fluctuation-dissipation relations for quantum systems with dissipation (atoms in radiation fields included), fluctuation processes in lasers, and nonequilibrium phase transitions in quantum systems. A number of basic problems of theoretical physics, such as superconductivity, superfluidity, flicker noise, quantum Hall-effect, were among his scientific interests. 
\begin{figure}
    \centering
    \includegraphics[width=0.5\textwidth]{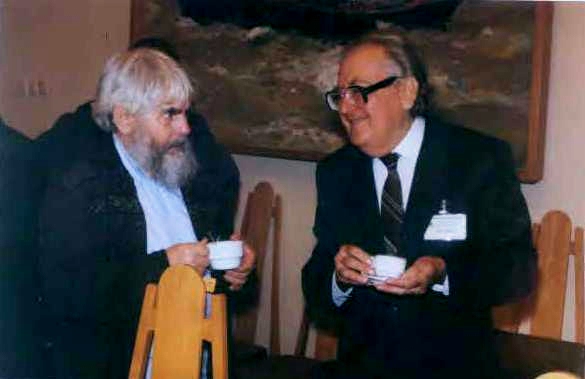}
    \caption{Yuri L.~Klimontovich with A. Sitenko during the Bogolyubov Conference ``Problems of Theoretical and Mathematical Physics'', Kyiv 1999. Photo by S. Filipchuk}
    \label{fig:klim99}
\end{figure}
Yu.L.~Klimontovich was a bright person possessing broad knowledge in diverse fields of science and culture. It was very interesting to discuss with him problems of physics, history, philosophy, literature and arts. Everybody who contacted him felt his friendly spirit. The scientific activity of Yu.L.Klimontovich has been honored in many countries. He won the State Prize of Russia (1991), the Sinel’nikov Prize of the Ukrainian Academy of Sciences (1990), the Kapitsa Gold Medal of the Russian Academy of Natural Sciences (1997), and the Prize of the Humboldt Foundation (1994). He was honored by the degree ``Doctor Philosophy Honoris Causa'' from Rostock University (1992), by a Soros professorship (1995, 1997) and a Senior Fellowship of the Cariplo Foundation (Italy, 1997). However,  the most important honor for Yu.L.~Klimontovich is the broad recognition of his scientific results by the international physics community. 
Many of his  works stimulated a lot of research activities, and they are still today highly interesting and inspiring.





\end{document}